\documentclass{ws-procs9x6}
\def\lan{\langle}
\def\ran{\rangle}
\def\beq{\begin{equation}}
\def\eeq{\end{equation}}
\def\bea{\begin{eqnarray}}
\def\eea{\end{eqnarray}}
\def\nn{\nonumber}
\def\al{\alpha}

\begin{document}

\title{Equation of state of quark-nuclear matter}

\author{G. Krein and V.E. Vizcarra}

\address{Instituto de F\'{\i}sica Te\'orica, Universidade Estadual Paulista \\
Rua Pamplona, 145, 01405-900 S\~ao Paulo, SP, Brazil \\
E-mail: gkrein@ift.unesp.br - indhy@ift.unesp.br}

\maketitle

\abstracts{Quark-nuclear matter (QNM) is a many-body system containing 
hadrons and deconfined quarks. Starting from a microscopic quark-meson coupling
(QMC) Hamiltonian with a density dependent quark-quark interaction, an 
effective quark-hadron Hamiltonian is constructed via a mapping procedure. 
The mapping is implemented with a unitary operator such that composites
are redescribed by elementary-particle field operators that satisfy canonical 
commutation relations in an extended Fock space. Application of the unitary 
operator to the microscopic Hamiltonian leads to effective, hermitian operators 
that have a clear physical interpretation. At~sufficiently high densities, 
the effective Hamiltonian contains interactions that lead to quark deconfinement. 
The equation of state of QNM is obtained using standard many-body techniques 
with the effective quark-hadron Hamiltonian. At~low densities, the model is 
equivalent to a QMC model with confined quarks. Beyond a~critical density,
when quarks start to deconfine, the equation of state predicted for QNM is
softer than the QMC equation of state with confined quarks.
}

\section{Introduction}
One of the most exciting open questions in the study of high density 
hadronic matter is the identification of the appropriate degrees of 
freedom to describe the different matter phases. For systems with 
matter densities several orders of magnitude larger than the nuclear 
saturation density, one expects a phase of deconfined matter composed of 
quarks and gluons whose properties very likely can be described 
by perturbative QCD. For ground state nuclei, there is a large body of 
experimental evidence that their gross properties can be described more 
economically employing hadronic degrees of freedom, rather than quarks and 
gluons. On the other hand, for matter at densities not asymptotically 
higher than the saturation density, like the ones in dense stars and 
produced in high-energy nuclear collisions, the situation seems to be very 
complicated, since hadrons and deconfined quarks and gluons can be simultaneously 
present in the system. Presently, it is not possible to employ QCD directly 
to study such systems and the use of effective, tractable models are essential for 
making progress in the field. 

One attractive model to study the different phases of hadronic matter in terms 
of explicit quark-gluon degrees of freedom is the quark-meson coupling (QMC) 
model, originally proposed by Guichon and subsequently improved by Saito 
and Thomas~\cite{guichon-ST}. For a list of references on further improvements 
of the model and recent work, see Ref.~\cite{exclvol}. In the QMC model, 
matter at low density is described as a system of nonoverlapping MIT bags 
interacting through effective scalar- and vector-meson degrees of freedom. 
The effective mesonic degrees of freedom couple directly to the quarks in the 
interior of the baryons. At very high density and/or temperature, when one 
expects that baryons and mesons dissolve, the entire system of quarks and gluons 
becomes confined within a single, big MIT bag. 

In a regime of very high density, the description of hadronic matter in terms 
of nonoverlapping bags should of course break down, since once the relative 
distance between two bags becomes much smaller than the diameter of a bag, 
the individual bags loose their identity. The density for which this starts 
to happen is presently unknown within QCD. 

In the present communication we introduce a generalization of the QMC model
that allows to include quark deconfinement at high density. Our starting point 
is a relativistic quark potential model~\cite{toki}. From the model quark 
Hamiltonian, we construct a unitarily equivalent Hamiltonian that contains 
quark and hadron degrees of freedom. Starting from the Fock-space representation 
of single-hadron states, a unitary transformation is constructed such that the 
composite-hadron field operators are redescribed in terms of elementary-particle 
field operators in an extended Fock space. When the unitary transformation is 
applied to the quark Hamiltonian, effective, hermitian Hamiltonians with a clear 
physical interpretation are obtained~\cite{FT}. In particular, one of such 
effective Hamiltonians describes the deconfinement of quarks from the interior 
of the hadrons. The equation of state of QNM can be calculated using standard 
many-body techniques with the quark-hadron Hamiltonian. We~will
show that at low densities, the model is equivalent to the QMC model and, 
beyond a critical density, when quarks start to deconfine, the equation of 
state predicted for QNM is softer than the QMC equation of state. 

\section{QMC model with confined quarks}

The nucleons are bound states of three constituent quarks. Constant scalar 
($\sigma_0$) and vector ($\omega_0$) meson fields couple to the constituent 
quarks in the interior of the nucleons. Each constituent quark satisfies a Dirac 
equation of the form
\beq
\left[ - i \vec{\alpha}\cdot{\vec\nabla} + \beta^0 m^*_q 
+ 1/2(1+\beta)V(r)\right]
\psi(r) = E^*_q \psi(r) ,
\label{dirac}
\eeq
where 
\bea
&& m^*_q = m_q - g^q_\sigma \, \sigma_0, \hspace{1.0cm} 
E^*_q = \varepsilon^* - g^q_\omega \, \omega_0 , \hspace{1.0cm}
V(r) = \sigma \, r. 
\label{fields}
\eea
The only difference with the model of Toki et al.~\cite{toki} is the form of
the potential, while theirs is a harmonic oscillator, ours is a linearly 
rising one. For a linearly rising potential, the Dirac equation cannot 
be solved analytically. We use the saddle point variational principle 
(SPVP)~\cite{saddle} to obtain an approximated solution. Since the Dirac 
Hamiltonian does not have a lower bound for the energy, the traditional
nonrelativistic variational method cannot be employed. The SPVP amounts to
 minimizing (maximizing) the energy expectation value with 
respect to the variational parameters corresponding to the upper (lower) 
component of the Dirac wave function. We use as ansatz for the Dirac wave
function~\cite{saddle}
\beq
\psi(r) = \left(\begin{array}{c}
u(r) \\
\displaystyle{
i \vec\sigma\cdot\hat{r} \, v(r) }
\end{array}  \right) \chi_s ,
\label{qspinor}
\eeq
with
\beq
u(r) = N e^{-\lambda^2 r^2/2}, \hspace{1.0cm}
v(r) =  i \, \gamma/\lambda \, \vec\sigma\cdot\vec\nabla \,u(r),
\label{ansatz}
\eeq 
where $N$ is a normalization constant, and $\lambda$ and $\gamma$ are the
variational parameters. The parameters are found by minimizing the energy 
eigenvalue $\varepsilon$ with respect to $\lambda$ and maximizing it with
respect to $\gamma$. We note here that for a harmonic oscillator potential,
the SPVP with this ansatz leads to the exact solution.

Following the traditional path in the QMC model, we initially fix the parameters
of the model in vacuum. The nucleon mass in vacuum ($\sigma_0 = 0 = \omega_0$) 
is given by
\beq
M_N = 3 \, \varepsilon - \varepsilon_0 \, \lambda  ,
\label{vacM}
\eeq
where the last term above is used to take into account the $c.m.$ energy of
the three-quark state and other short-distance effects not taken into account 
by the confining potential, such as gluon exchange. $\varepsilon_0$ is 
fitted to obtain $M_N = 939$~MeV. The value of the string tension is taken 
to be $\sigma = 0.203$~GeV$^2$ and $m_q = 313$~MeV.  With these parameters, 
the SPVP leads to $\lambda = 2.38$~fm$^{-1}$ and $\gamma = 0.346$. The value 
required for $\varepsilon_0$ to fit the nucleon mass is $4.67$~MeV~fm. 

Next, we proceed to obtain the energy of nuclear matter. Nuclear matter in
the QMC model is modeled as a system of nucleons treated in the mean field
approximation, in which the quarks remain confined within the nucleons.  
The nucleon mass is now obtained as above, but now including the mean fields 
coupled to the quarks. The energy density of symmetrical nuclear matter is 
given by the traditional expression in the QMC model
\beq
\frac{E}{V} = 4 \int^{k_F}_0 \frac{d^3 p}{(2\pi)^3}\, 
E^*_N(p)\, 
+ 3\,g^q_{\omega}\omega_0\,\rho_B +  \frac{1}{2}m^2_{\sigma}\sigma^2_0
- \frac{1}{2} m^2_{\omega}\omega^2_0 ,
\label{EMF} 
\eeq
where $E^*_N = \sqrt{p^2 + M^{* 2}_N}$, $\rho_B$ is the baryon density and
$m_{\sigma}$ and $m_\omega$ are the masses of the mesonic excitations. The
next step consists in determining the mean fields. The vector mean field,
from its equation of motion, is simply given in terms of $\rho_B$, and 
the scalar field is obtained by minimizing $E$ with respect to $\sigma_0$, 
as usual. The coupling constants are then obtained by fitting $E$ to the 
binding energy of nuclear matter at the saturation density, i.e.  
$E/B - M_N = - 15.7$~MeV at $\rho_B = \rho_0 = 0.17$~fm$^{-3}$ 
(or $k_F = 1.36$~fm$^{-1}$), where $B$ is the baryon number - in this case
$B$ is equal do the number of nucleons. Note that for each value of $\rho_B$, 
one has to use the SPVP to obtain the in-medium values of $\lambda$ and 
$\gamma$. The values obtained for the coupling constants are $g^q_\sigma 
= 6.1355$ and $ 3 g^q_\omega = 6.285$. The incompressibility is found to be 
$K = 248$~MeV.

In the next section we generalize the model to allow the deconfinement of
quarks. 

\section{QMC model with quark deconfinement}

Here we construct an effective Hamiltonian that contains hadron and quark 
degrees of freedom. The starting point is the quark model discussed in the
previous section. In this model, the one-nucleon state can be written in a 
second-quantized notation 
as
\beq
|\alpha\ran = B^{\dag}_{\alpha}|0 \ran, \hspace{1.0cm}
B^{\dag}_{\al}=\frac{1}{\sqrt{3!}} 
\Psi_{\alpha}^{\mu_1\mu_2\mu_3}
q_{\mu_1}^{\dag}q_{\mu_2}^{\dag}q_{\mu_3}^{\dag},
\label{nucleon}
\eeq
where the $q_{\mu}^{\dag}$'s are constituent-quark creation operators and 
$\Psi_{\alpha}^{\mu_1\mu_2\mu_3}$ is the Fock-space nucleon amplitude - for
independent quarks, this is simply the product of three single-quark wave 
functions. The convention of summing over repeated indices is used throughly. 
The quark creation and annihilation operators satisfy the usual 
canonical anticommutation relations
\beq
\{q_{\mu}, q^{\dag}_{\nu}\} = \delta_{\mu \nu},\hspace{1.0cm}
\{q_{\mu}, q_{\nu}\} = 
\{q^{\dagger}_{\mu}, q^{\dagger}_{\nu}\} = 0.
\label{qcom}
\eeq
The index $\alpha$ denotes the spatial and internal quantum numbers, such as
internal and c.m. energies and the spin-isospin quantum numbers of the nucleon. 
Similarly, the quark indices $\mu$ identify the spatial and internal quantum
numbers as momentum, spin, flavor and color. The amplitude 
$\Psi_{\alpha}^{\mu_1\mu_2\mu_3}$ is taken to be orthonormalized:
\beq
\lan  \alpha|\beta \ran= \Psi_{\alpha}^{*\mu_1\mu_2\mu_3 } 
\Psi_{\beta}^{\mu_1\mu_2\mu_3}=\delta_{\alpha \beta}\;.
\label{norm}
\eeq

In the abbreviated notation we are using, the Hamiltonian corresponding to 
Eq.~(\ref{dirac}) can be written as
\beq
H_q =  T_q + V_{qq} = T\left(\mu\right)q^{\dag}_{\mu} q_{\mu} + 
\frac{1}{2} V_{qq}\left(\mu\nu;\sigma\rho\right)
q^{\dag}_{\mu}q^{\dag}_{\nu}q_{\rho}q_{\sigma},  
\label{qHamilt}
\eeq
where $V_{qq}$ is the confining potential. 

Using the quark anticommutation relations of Eq.~(\ref{qcom}) and the 
normalization condition of Eq.~(\ref{norm}), one can shown that the nucleon 
operators, $B_\alpha$ and $B^\dag_\alpha$, satisfy the following 
anticommutation relations 
\beq
\{B_{\alpha},B_{\beta}^{\dag}\}  = \delta_{\alpha\beta}-\Delta_{\alpha\beta},
\hspace{1.0cm} \{B_{\alpha},B_{\beta}\} = 0 ,
\label{ft6}
\eeq
where
\bea
\Delta_{\alpha\beta}=3\Psi_{\alpha}^{*\mu_1\mu_2\mu_3}\Psi_{\beta}^
{\mu_1\mu_2\nu_3}q_{\nu_3}^{\dag}q_{\mu_3}-\frac{3}{2}\Psi_{\alpha}^
{*\mu_1\mu_2\mu_3}\Psi_{\beta}^{\mu_1\nu_2\nu_3}q_{\nu_3}^{\dag}q_{\nu_2}^
{\dag}q_{\mu_2}q_{\mu_3}. 
\label{ft7}
\eea
In addition, one has
\bea
\{q_{\mu},B_{\alpha}^{\dag}\} = \sqrt{ \frac{3}{2} }
\Psi_{\alpha}^{\mu\mu_2\mu_3}q_{\mu_2}^{\dag}q_{\mu_3}^{\dag},
\hspace{1.5cm}
\{q_{\mu},B_{\alpha}\}=0.
\label{ft8}
\eea
The term $\Delta_{\alpha \beta}$ is responsible for the noncanonical nature 
of the baryon anticommutator. This term and the nonzero value of Eq.~(\ref{ft8}) 
are manifestations of the composite nature of the baryons and the kinematical 
dependence of the quark operator and nucleon operators. This fact complicates
enormously the mathematical treatment of many-body systems in which deconfined 
quarks and nucleons are simultaneously present. The mapping formalism of
Ref.~\cite{FT}, known as the Fock-Tani (FT) representation~\cite{girar}, is a
way to circumvent such complications. We will shortly review this formalism
in the context of the present model. For further details, and applications 
for more general models, the reader is referred to Ref.~\cite{FT}.

The basic idea is to extend the original Fock space by introducing fictitious,
or {\em ideal} nucleons that satisfy canonical anticommutation relations. The
unitary operator is constructed in the extended Fock space such that
\bea
|\alpha\ran=B^{\dag}_{\alpha}|0 \ran
\longrightarrow U^{-1}|\alpha \ran\equiv |\alpha)=
b^{\dagger}_{\alpha}|0),
\label{single_bar}
\eea
where ideal baryon operators $b^{\dagger}_{\alpha}$ and $b_{\alpha}$ satisfy,
by definition, canonical anticommutation relations
\bea
\{b_{\alpha}, b^{\dagger}_{\beta}\}=\delta_{\alpha \beta},
\hspace{1.0cm}
\{b_{\alpha}, b_{\beta}\} = 0 .
\label{banticom}
\eea
The state $|0)$ is the vacuum of both $q$ and $b$ degrees of freedom in the new
representation.
In addition, in the new representation, the quark operators $q^{\dagger}$ and
$q$ are kinematically independent of the $b^{\dagger}_{\alpha}$ and 
$b_{\alpha}$
\bea
\{q_{\mu},b_{\alpha}\}=
\{q_{\mu},b^{\dagger}_{\alpha}\}=0\:.
\label{indep_bar}
\eea
The unitary operator $U$ can be constructed as a power series in the
bound state amplitude $\Psi$. The rational for this is clear: in situations 
that the quarks remain confined in the interior of the nucleons, the term
$\Delta_{\alpha\beta}$ plays no role, can be taken to be zero and the
unitary operator becomes trivial~\cite{girar}~\cite{FT}. This is
the situation for low densities, when the internal structures of the 
nucleons do not overlap significantly in the system. As the density of 
the system increases, the quark structures of different nucleons start 
to overlap. An expansion in powers of $\Psi$'s offers a power counting
procedure to construct the unitary operator. 

The effective Hamiltonian is constructed by applying the unitary operator 
to the microscopic quark Hamiltonian of Eq.~(\ref{qHamilt}), 
$H_{eff} = U^{-1}H_qU$. The zeroth-order $U$ is trivial
and not interesting. The first-order $U$ brings interesting effects. 
At this order, we denote the effective Hamiltonian by $H^{(1)}_{eff}$, where 
the superscript~$(1)$ means that $U$ has been evaluated up to the first order 
in $\Psi$. $H^{(1)}_{eff}$ can be written as
\beq
H^{(1)}_{eff} = T_q + H_b + \tilde V_{qq} +  V_{qb} + \cdots \,.
\label{Heff1}
\eeq
The $\cdots$ refer to terms not relevant for our discussion here. $H_b = 
T_b + V_{bb}$, where $T_b$ is a single-nucleon energy and $V_{bb}$ is an
effective nucleon-nucleon interaction without quark exchange. This term 
leads to the normal QMC model, in which the many-body system is described by 
nonoverlapping nucleons - no quark-exchange. In particular, it can describe 
Fock terms in the QMC model~\cite{fock}. The term $\tilde V_{qq}$ contains
two-quark and three-quark interactions. It can be shown that if $\Psi$ is 
a bound-state eigenstate of the original quark Hamiltonian, $\tilde V_{qq}$
collapses to
\bea
V_{qq}&=& \frac{1}{2} \,
V_{qq}(\mu\nu;\sigma\rho)\,
q^{\dag}_{\mu}q^{\dag}_{\nu}q_{\rho}q_{\sigma}
- {E}_{\alpha} B^{\dag}_{\alpha} B_{\alpha}.
\label{ft25_4}
\eea
It is not difficult to show that this interaction leads to a quark Hamiltonian
that has a positive semidefinite spectrum. That is, after the transformation, 
the resulting Hamiltonian involving only quark operators is unable to bind 
three quarks to form a nucleon, it describes only states in the continuum. 
The term $V_{qb}$ is given by
\bea
V_{bq} &=&\frac{1}{\sqrt{6}} \Bigl[
H(\mu_1\mu_2;\sigma\rho) \Psi^{\sigma\rho\mu_3}_{\beta} \nn\\
&-& H(\mu\nu;\sigma\rho)\Psi^{\sigma\rho\tau_3}_{\beta}
\Delta(\mu_1\mu_2\mu_3;\mu\nu\tau_3)
\Bigr]
q^{\dag}_{\mu_1}q^{\dag}_{\mu_2}q^{\dag}_{\mu_3}b_{\beta} + {\rm h.c.}\,,
\label{bq}
\eea
where $h.c.$ denotes hermitian conjugation and 
$\Delta (\mu \nu\tau;\sigma \rho \lambda ) = \sum_\alpha \Psi^{\mu \nu\tau}_{\al} 
\Psi^{\ast\sigma\rho\lambda}_{\al}$ is known as the bound-state kernel.
If $\Psi$ is a stationary state of the microscopic quark Hamiltonian, one
obtains 
\beq
V_{bq} = 0,
\label{nobreak}
\eeq
since $\Delta (\mu \nu\tau;\sigma \rho \lambda )\Psi^{\sigma \rho\lambda}_{\al}
= \Psi^{\mu \nu \tau}_{\al}$.  This result reflects the stability
of the baryon bound state to spontaneous decay in the absence of external 
perturbations. This is clearly the case for a nucleon in vacuum. Also, in the QMC
model, when one constructs the unitary transformation $U$ with a $\Psi$ that
is an eigenstate of the microscopic quark Hamiltonian with the mean fields 
$\sigma_0$ and $\omega_0$ in it, the term $V_{bq}$ continues to be zero, and
explicit quark degrees of freedom will not be present in the system at this
order of $\Psi$. 

Now, in a many-body system, the confining quark-quark interaction will become
modified due to a variety of effects. Some of such effects, as self-energy
corrections from quark loops, can be calculated within the model using 
standard many-body techniques. However, in a high density system there are 
other QCD effects that are not captured by the model, such as pair creation 
and gluonic interactions, that eventually will lead to quark deconfinement. 

The formalism we just described allows to include in an effective way 
deconfinement in the QMC model. One generates an effective quark-hadron 
Hamiltonian as above using $\Psi$'s that are eigenstates of the QMC Hamiltonian, 
Eq.~(\ref{dirac}). Now, if $V(r)$ is modified such as that it does not lead to 
absolute confinement, the term $V_{bq}$, given by Eq.~(\ref{bq}), is not zero. 
In a mean field perspective, the Hamiltonian of Eq.~(\ref{Heff1}) leads to
two Fermi seas, one for baryons and one for quarks. The crucial, and difficult
point here is to obtain the relative abundances of baryons and quarks in the 
system. This can be evaluated in an approximated way as follows.

Let $Z$ be the fraction of baryons in the system, 
\beq
\sum_{\alpha} \lan b^{\dagger}_{\alpha} b_{\alpha}\ran = Z B,\hspace{1.0cm}
\sum_{\mu} \lan q^{\dagger}_{\mu} q_{\mu}\ran = (1-Z) B,
\eeq
where $B$ is as in the previous section the total baryon number. In the mean 
field approximation - or independent-particle approximation - and for 
sufficiently small $V_{bq}$, $Z$ can be estimated by the perturbative formula
\beq
Z^{-1} = 1 + (b|V^{\dagger}_{bq}\frac{P}{H_0 - E_0}V_{bq}|b) ,
\label{Z-1}
\eeq
where $H_0$ is $T_q + T_b$, and $P = 1 - |b)(b|$ is a projection operator.

\begin{figure}[ht]
\vspace{-5.0cm}
\centerline{\epsfxsize=4.2in\epsfbox{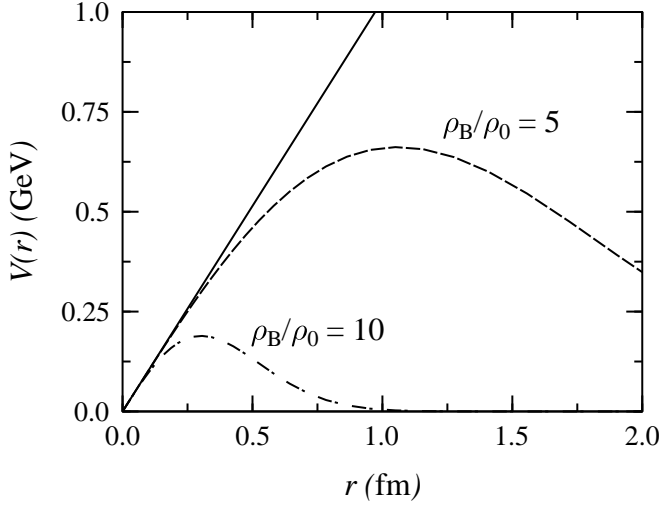}}   
\vspace{-1.2cm}
\caption{The confining potential in vacuum (solid line) and in matter for
two different baryon densities.
\label{pot}
}
\end{figure}

In order to evaluate Eq.~(\ref{Z-1}), we postulate a density dependence for 
the confining interaction of the form~\cite{walb}
\beq
V(r) = \sigma \, r \,e^{- \mu^2 \,r^2},
\label{densV}
\eeq
where $\mu$ is a prescribed function of $\rho_B$. We use a simple formula for
$\mu$, such that it is zero for baryon densities below three times the normal 
nuclear matter density $\rho_0$, and for higher densities it increases linearly 
with the density as $\mu = \rho_B/3\rho_0 - 1$. In Fig.~(\ref{pot}) we show the 
potential of Eq.~(\ref{densV}) for zero density, and 5 and 10 times the 
saturation density of normal nuclear matter. 

The energy density of the system can be written as
\bea
\frac{E}{V} &=& 4 \int^{k^b_F}_0 \frac{d^3 p}{(2\pi)^3}\, 
E^*_N(p)\, + 3\,g^q_{\omega}\omega_0\,\rho_B 
+  \frac{1}{2}m^2_{\sigma}\sigma^2_0 - \frac{1}{2} m^2_{\omega}\omega^2_0 \nn\\
&+& 12 \int^{k^q_F}_0 \frac{d^3 k}{(2\pi)^3}\, E^*_q(k) ,  
\label{EMFq} 
\eea
where $E^*_q(k) = \sqrt{k^2 + m^{*2}_q}$ and the Fermi momenta $k^b_F$ and $k^q_F$ 
are related to the nucleon density and quark density as
\beq
\rho_b = Z \rho_B = 2 (k^b_F)^2 /3\pi^2, \hspace{1.0cm} 
\rho_q = (1-Z) \rho_B =  2 (k^q_F)^2/ \pi^2, 
\eeq
At this point, it is important to notice that since $\mu$ only starts to operate
for densities larger than three times the normal density, the coupling constants 
$g^q_\sigma$ and $g^q_\omega$ are the same as before. Of course, for higher
densities, there is a somewhat complicated self-consistency problem to be
solved, since $Z$ is density dependent. Therefore, in the process of obtaining
$\sigma$, the iterative problem becomes more complicated.

\begin{figure}[ht]
\vspace{-5.0cm}
\centerline{\epsfxsize=4.2in\epsfbox{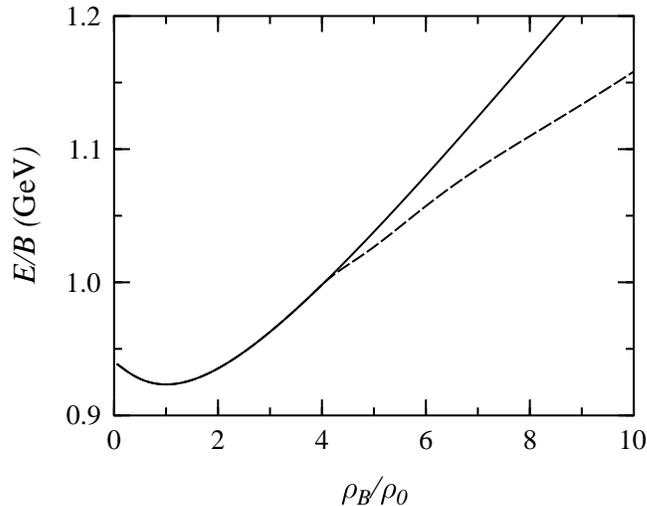}}   
\vspace{-1.2cm}
\caption{Equation of state of quark nuclear matter. The solid line is 
for matter composed of nucleons only and the dashed line is for matter
composed by nucleons and quarks.
\label{eqnm}
}
\end{figure}

In order to proceed, we need $Z$ as a function of $\rho_B$. It can be calculated
numerically with our ansatz wave function given above. The calculation, however,
involves multidimensional integrals that must be done using a Monte Carlo 
integrator. For our purposes here, in this initial investigation we make some
approximations. Initially we neglect the lower component of the Dirac spinor.
This does not seem to be a too drastic approximation, since $\gamma$ in
Eq.~(\ref{ansatz}) is a small quantity. In this approximation, one obtains 
\beq
Z^{-1} =  1 + \int d^3 k_1 d^3 k_2 d^3 k_3\, |\Phi_p(k_1,k_2,k_3)|^2
\frac{|F(\vec k_1 - \vec k_2)|^2}{\Delta E(p,k_1,k_2,k_3)},
\label{explZ}
\eeq
with $F(\vec q)$ given by
\beq
F(\vec q) = \int d^3 k \, \Delta \tilde V(\vec k) \, e^{ - k^2/\lambda^2}
\Bigl(e^{ \, \vec k\cdot \vec q/\lambda^2} - 1\Bigr),
\label{F}
\eeq
where $\Phi_p(k_1,k_2,k_3)$ is the three-quark wave function of the nucleon
with c.m. momentum $\vec p$ (see Ref.~\cite{FT}), $\Delta E(p,k_1,k_2,k_3)$ 
is the difference between of the energies of the three unbound quarks and of 
the three quarks bound in the potential, and $\Delta \tilde V(\vec k)$ is the 
Fourier transform of $\Delta V(r)$, where
\beq
\Delta V(r) = \sigma\, r \, \Bigl( e^{- \mu^2 \,r^2} - 1\Bigr).
\eeq
This clearly shows that once $\mu=0$, i.e. the potential is density independent,
one regains the original QMC model.

Now, Eqs.~(\ref{explZ}) and (\ref{F}) still require a lot of numerical
work. We simplify them further by making two additional approximations. The
first one consists in neglecting the momentum dependence of the energy 
denominator and the second one is to use an average value for $q^2$ in 
$F(q^2)$. Both approximations taken together seem not to be a bad approximation,
since the energy denominator under the integral is dominated by low momenta.
Now the problem consists in a single one dimensional integral that can easily
be performed with a Gauss integration. 

In Fig.~(\ref{eqnm}) we present the results for the energy per baryon number, 
$E/B$ as a function of the ratio $\rho_B/\rho_0$. The solid line in this figure 
is the result for the QMC model of the previous section. The dashed line shows 
that the deconfining of quarks leads to a softening of the equation of state.
It would be interesting to investigate the consequences of this softening for
neutron-star phenomenology. Soft equations of state seem to be required to 
explain recent observational data of compact stellar objects.

\section{Conclusions}

We have generalized the QMC model to include quark deconfinement in matter.
The model is based on an effective quark-hadron Hamiltonian obtained
via a mapping procedure from a relativistic microscopic quark Hamiltonian 
with a density dependent quark-quark interaction. The equation of state of 
QNM was obtained using the effective quark-hadron Hamiltonian. It was found 
that beyond a critical density, when quarks start to deconfine, the equation 
of state predicted for QNM is softer than the usual QMC equation of state. 
Implications of this equation of state for the phenomenology of compact stellar 
objects were pointed out.

\section*{Acknowledgments}
Work partially supported by the Brazilian agencies CNPq and FAPESP.


\begin{thebibliography}{0}
\bibitem{guichon-ST} P. A. M. Guichon, {\it Phys. Lett.} {\bf B200}, 235 
(1988); K. Saito and A.W. Thomas, {\it Phys. Lett.} {\bf B327}, 9 (1994).

\bibitem{exclvol} P.K. Panda, M.E. Bracco, M. Chiapparini, E. Conte and G.Krein,
{\it Phys. Rev. C} (in press), nucl-th/0205051. 
\bibitem{toki} H. Toki, U. Meyer, A.~Faessler, and R. Brokmann, 
{\it Phys. Rev.} {\bf C58}, 3749 (1998).

\bibitem{FT} D. Hadjimichef, G. Krein, S. Szpigel and J.S. da Veiga,
{\em Phys. Lett.} {\bf B367}, 317 (1996); {\it Ann. Phys. (NY)}, {\bf 268}, 
105 (1998).

\bibitem{saddle} J. Franklin and R.L. Intemann, {\em Phys. Rev. Lett.}
{\bf 54}, 2068 (1985); J.D. Talman, {\em Phys. Rev. Lett.} {\bf 57}, 
1091 (1986).

\bibitem{girar} M.D. Girardeau, {\it Phys. Rev. Lett.} {\bf 27}, 1416 (1971);
{\it J. Math. Phys.} {\bf 16}, 1901 (1975).

\bibitem{fock} G. Krein, A.W. Thomas and K. Tsushima, {\it Nucl. Phys.} 
{\bf A650}, 313 (1999).

\bibitem{walb} W. Alberico, P. Czerski and M. Nardi, {\em Eur. Phys. J.} {\bf A4}, 195 (1999).

\end{thebibliography}
\end{document}